\documentclass{article}

\newcommand{\C}{{\bf C}}

\newcommand{\N}{{\bf N}}
\newcommand{\R}{{\bf R}}
\def\kasten{$~~\mbox{\hfil\vrule height6pt width5pt depth-1pt}$ }

\newtheorem{theorem}{Theorem}
\title{On the formulation of SPDEs leading to local, relativistic
 QFTs with indefinite metric and nontrivial S-matrix}
\author{Sergio Albeverio and Hannpo Gottschalk\\ Institut f\"ur angewandte Mathematik\\
Rheinische Fridrich-Wilhelms-Universit\"at Bonn,\\
D-53155 Bonn, Germany\\ albeverio@uni-bonn.de/gottscha@wiener.iam.uni-bonn.de\\ 
Jiang-Lun Wu\\Fakult\"at f\"ur Mathematik, Ruhr-Universit\"at Bochum,\\
 D-44780 Bochum, Germany\\jiang-lun.wu@ruhr-uni-bochum.de}

\begin{document}
\maketitle
\begin{abstract}
We discuss Euclidean covariant vector random fields as the
solution of stochastic partial differential equations of the form
$DA=\eta$, where $D$ is a covariant (w.r.t. a representation
$\tau$ of $SO(d)$) differential operator with "positive mass
spectrum" and $\eta$ is a non-Gaussian white noise. We obtain
explicit formulae for the Fourier transformed truncated Wightman
functions, using the analytic continuation of Schwinger functions
discussed by Becker, Gielerak and {\L}ugewicz. Based on these
formulae we give necessary and sufficient conditions on the mass
spectrum of $D$ which imply nontrivial scattering behaviour of
relativistic quantum vector fields associated to the given
sequence of Wightman functions. We compute the scattering
amplitudes explicitly and we find that the masses of particles in
the obtained theory are determined by the mass spectrum of $D$.
\end{abstract}

\section{Introduction}
Local and relativistic quantum fields can be obtained via analytic
continuation from Euclidean random fields.
From the 80-ies on, this concept, which proved to be useful especially
for scalar fields in the space-time dimension $d=2$, was applied in
a number of articles (see e.g. \cite{AIK} and references therein) to
random vector fields obtained as solutions of the stochastic partial
differential
equation (SPDE) $DA=\eta$,
where $D$ is a complex, quaternionic or octonionic Cauchy-Riemann
differential operator and $\eta$ a non Gaussian noise with values
in the fields of complex numbers, quaternions or octonions,
respectively. For the quaternionic case (space-time dimension $d=4$)
we have recently been able to prove that the associated quantum
gauge fields have non-trivial scattering behaviour \cite{AGW3}.

In the paper \cite{BGL}, the equation $DA=\eta$ was studied
systematically in $d$ dimensional space-time and spin dimension $L$
(of $\eta$).  Let $\tau:SO(d)\to Gl(L)$ be a
representation s.t. $\eta$ transforms covariantly under the
representation
of the Euclidean group induced by $\tau$, i.e.
$\tau(\Lambda)\eta(\Lambda^{-1}(x-y))=\eta(x)$ in probability
law $\forall \Lambda\in SO(d)$ and $y\in \R^d$. Let
$D$ be a differential operator covariant w.r.t. the representation
$\tau$, i.e. $\tau(\Lambda) D_x\tau(\Lambda^{-1})=D_{\Lambda x}
\forall \Lambda \in
SO(d)$. Furthermore we assume that $D$ is translation invariant,
i.e. has constant coefficients. In this case, for the Fourier
transformed Green's function $\hat D^{-1}(k)$ of $D$ the following
representation was obtained in \cite{BGL}:
\begin{equation}
\label{1eqa}
\hat D^{-1}(k)= {Q_E(k)\over \prod_{l=1}^N(|k|^2+m_l^2)^{\nu_l}}
\end{equation}
with $m_l\in\C,m_j\not=m_l$ for $ l\not = j$ and $\nu_l\in \N$.
$Q_E(k)$ is an $L\times L$-matrix with polynomial entries of order
$\leq \kappa=2(\sum_{l=1}^N\nu_l-1)$ which fulfills the Euclidean
transformation law $\tau(\Lambda) Q_E(k)\tau(\Lambda^{-1})=Q_E(\Lambda k)\forall
\Lambda \in SO(d)$. Without loss of generality we assume that
$Q_E$ is prime w.r.t the factors $(|k|^2-m_l^2)$, i.e. that none of
them divides all of the polynomial matrix elements of $Q_E$. If
one imposes  a "positive mass spectrum" condition $m_l>0$ for
$l=1,\ldots,N$ it
follows immediately that $D$ is invertible on the space of
($\C^L$-valued)  distributions over $\R^d$ and we can
therefore solve the above SPDE by setting $A=D^{-1}\ast\eta$.

The Schwinger functions (moments) associated to $A$ can be calculated
 exlicitly as
\begin{eqnarray}
\label{2eqa}
&\displaystyle S_{n,\alpha_1\cdots\alpha_n}(x_1,\ldots,x_n)=
{\bf E}\left[A_{
\alpha_1}(x_1)\cdots A_{\alpha_n}(x_n)\right]& \nonumber\\
&\displaystyle = \sum_{I\in{\cal P}^{(n)}}\prod_{\{j_1,\ldots,j_l\}
\in I}\underbrace{C^{\beta_{j_1}\cdots\beta_{j_l}}\int_{\R^d}
\prod_{r=1}^lD^{-1}_{\beta_{j_r}\alpha_{j_r}}(x_{j_r}-x)~dx}_{=S^T_{n,
\alpha_{j_1}\cdots\alpha_{j_l}}(x_{j_1},\ldots,x_{j_l})}&
\end{eqnarray}
where ${\cal P}^{(n)}$ is the collection of all partitions
of $\{1,\ldots,n\}$ into disjoint subsets and $C^{\beta_{j_1}
\cdots\beta_{j_l}} $ are constants depending on the law of $\eta$.
We have also used the Einstein convention of summation on repeated
upper and lower greek indices w.r.t. an arbitrary invariant metric
on the spin-space.

In this paper we  study the analytic continuation of the
truncated Schwinger functions $S_n^T$ to truncated relativistic
Wightman functions $W_n^T$ and the scattering behaviour of
these Wightman functions. The paper is organized as
follows: In Section 2 we explicitly compute the Fourier transformed
truncated Wightman functions of the model. This will be done
combining techniques of \cite{AGW1} and \cite{BGL}. In Section 3 we
discuss the scattering behaviour of the truncated Wightman
functions, showing that it is nontrivial, if and only if $\nu_l=1$
for $l=1,\ldots,N$. In this case we derive explicit formulae for the
truncated scattering amplitudes.

\section{Analytic continuaton of the Schwinger functions}

In this section we obtain a representation of the truncated Schwinger
functions $S_n^T$ as Fourier-Laplace transforms, i.e.
\begin{eqnarray}
\label{3eqa}
&\displaystyle S^T_{n,\alpha_1\cdots\alpha_n}(x_1,\ldots,x_n)={\cal L}
(\hat
W_{n,\alpha_1\cdots\alpha_n}^T)(x_1,\ldots,x_n)=& \nonumber \\
&\displaystyle (2\pi)^{-dn/2}\int_{\R^{dn}} \exp(\sum_{l=1}^n-k^0_l
x_l^0+i\vec k_l\cdot \vec x_l)
~ \hat W_{n,\alpha_1\cdots\alpha_n}^T(k_1,\ldots,k_l)
~dk_1\cdots dk_n ~,&
\end{eqnarray}
where $\hat W_{n\alpha_1\cdots\alpha_n}^T$ is a tempered distribution
which fulfils the spectral property, i.e. it has support in the cone
$\{(k_1,\ldots,k_n)\in \R^{dn}:
q_j=\sum_{l=1}^j k_l \in \bar V_0^-, j=1,\ldots,n-1\}$, and
$x_1^0<\ldots<x_n^0$. Here, $\bar V_0^-$ stands for the closed
backward lightcone (that
we do not use the forward lightcone for the formulation of the
spectral condition
is a matter of convention on the Fourier transform).
Under this condition the above integral exists
in the sense of tempered distributions. If such a representation
exists,
it follows from the general theory of quantum fields that $S_n^T$
is the
analytic continuation of the inverse Fourier transform $W_n^T={\cal F}^{-1}
(\hat W_n^T)$ of $\hat W_n^T$ from points with
purely relativistically real time to the Euclidean points of purely
imaginary time. Furthermore, it follows from the symmetry and Euclidean
covariance of the $S_n^T$ that $W_n^T$ fulfills the requirements of
Poincar\'e covariance and locality, see e.g. \cite{OS}.

In order to obtain such a representation, we apply the formula for the
expansion of an inverse polynomial into partial fractions on the inverse
of the denominator of Eq. (\ref{1eqa})
\begin{equation}
\label{4eqa}
{1\over
\prod_{l=1}^N(|k|^2+m_l^2)^{\nu_l}}=\sum_{l=1}^N\sum_{j=1}^{\nu_l}
{b_{lj}\over (|k|^2+m_l^2)^j}
\end{equation}
with $b_{lj}\in \R$ uniquely determined and $b_{l\nu_l}\not =0$. Thus,
$D^{-1}(x)$ can be represented as $Q_E(-i\nabla) \sum_{l=1}^N\sum_{j=1}^{\nu_l}
b_{lj}(-\Delta+m_l^2)^{-j}(x)$ and
\begin{equation}
\label{5eqa}
S_{n,\alpha_1\cdots\alpha_n}^T= {\bf Q}_{E,n}(-i\underline{\nabla})_{\alpha_1
\cdots\alpha_n}\sum_{l_1,\ldots,l_n=1}^N
\sum_{j_1,\ldots,j_n=1}^{\nu_1,\ldots,\nu_n} \prod_{r=1}^nb_{l_r,j_r}
S_{n,(m_{l_1},j_1),\ldots,
(m_{l_n},j_n)}
\end{equation}
with ${\bf Q}_{E,n}(-i\underline{\nabla})_{\alpha_1\cdots\alpha_n}=
C^{\beta_1\cdots\beta_n}Q_E(-i\nabla)^{\otimes n}_{\beta_1\cdots,
\beta_n
\alpha_1\cdots\alpha_n}$ and
\begin{equation}
\label{6eqa}
S_{n,(m_{l_1},j_1)\cdots
(m_{l_n},j_n)}(x_1,\ldots,x_n)=\int_{\R^d}\prod_{r=1}^n(-
\Delta+m_{l_r}^2)^{-j_r}(x_r-x)~dx
\end{equation}
In order to obtain a Laplace representation as in Eq. (\ref{3eqa}) for
$S_{n,(m_{l_1},j_1),\ldots,(m_{l_n},j_n)}$ we introduce some notations.
Let $\delta_m^{(j)\pm(k)}:=\theta (\pm k^0>0)\delta^{(j)}(k^2-m^2)$ where
$\delta^{(j)}$ is the $j$-th derivative of the one-dimensional delta
distribution, $\theta$ is the Heaviside function and $k^2={k^0}^2-|\vec
k|^2$. Furthermore, let
\begin{equation}
{\bf Q}_{M,n}(\underline{k})={\bf Q}_{M,n}((k_1^0,\vec k_1),\ldots,
 (k_n^0,\vec k_n))={\bf Q}_{E,n}((ik_1^0,\vec k_1),\ldots,(i k_n^0,
\vec k_n)).
\end{equation}
 We are now in the position to state the main theorem of this section:

\begin{theorem}{Theorem} (i) For $n\geq 3$ or $n=2, m_{l_1}\not=m_{l_2}$ let
 $\hat W_{n,(m_{l_1},j_1),\ldots , (m_{l_n},j_n)}^T$ be
defined as
\begin{eqnarray}
\label{7eqa}
(2\pi)^{-{d(n-2)+2\over 2}}\left\{ \sum_{s=1}^n\prod_{v=1}^{s-1}
 {(-1)^{j_v-1}\over (j_v-1)!}\delta_{m_{l_v}}^{-,(j_v-1)}
 (k_v){1\over (k_s^2-
m_{l_s}^2)^{j_s}}\right.&&\nonumber \\
\times \left. \prod _{v=s+1}^n{(-1)^{j_v-1}\over (j_v-1)!}
 \delta_{m_{l_v}}^{+,(j_v-1)}
(k_v)\right\}
\delta (\sum_{s=1}^n k_s)&&
\end{eqnarray}
and for $n=2, m_{l_1}=m_{l_2}$:
\begin{eqnarray}
\label{7aeqa}
&\displaystyle {(-1)\over (j_1+j_2-1)!}\Bigg\{\sum_{l=0}^{j_1+j_2-1}
{\scriptsize
\left(\begin{array}{c} j_1+j_2-1\\ l \end{array}\right)}{(-1)^l l!
\delta^{-,(n-l)}_{m_{l_1}}(k_1)\over4(|\vec k_1|^2+m_{l_1}^2)^l}
 \nonumber \\
&\quad\quad\quad\quad\quad\quad\displaystyle +(-1)^{j_1+j_2-1}
\delta_{m_{l_1}}^{-,(j_1+j_2-1)}(k_1)\Bigg\}\delta(k_1+k_2)\, .
\end{eqnarray}
Then $\hat W^T_{n,(m_{l_1},j_1),\ldots,(m_{l_n},j_n)}$ is a tempered
distribution on $\R^{dn}$ which fulfills the spectral property and
\begin{equation}
\label{7beqa}
S^T_{n,(m_{l_1},j_1),\ldots,(m_{l_n},j_n)}={\cal L}(\hat
W_{n,(m_{l_1},j_1),\ldots,(m_{l_n},j_n)}^T).
\end{equation}

\noindent (ii) Let $\hat W_{n,\alpha_1\cdots\alpha_n}^T$ be defined as
\begin{equation}
\label{8eqa}
\hat W_{n,\alpha_1\cdots\alpha_n}^T={\bf Q}_{M,n}
(\underline{k})_{\alpha_1\cdots,\alpha_n}\sum_{l_1,\ldots,l_n=1}^N
\sum_{j_1,\ldots,j_n=1}^{\nu_1,\ldots,\nu_n} \prod_{r=1}^nb_{l_rj_r}
 \hat W^T_{n,(m_{l_1},j_1),\ldots,
(m_{l_n},j_n)},
\end{equation}
then $S_{n,\alpha_1\cdots,\alpha_n}^T={\cal L}(\hat
W_{n,\alpha_1\cdots,\alpha_n}^T)$. Furthermore, $
W_{n,\alpha_1\cdots,\alpha_n}^T={\cal F}^{-
1}(\hat W_{n,\alpha_1\cdots,\alpha_n}^T) $ fulfills the requirements of
temperedness, relativistic covariance w.r.t. the representation $\tilde \tau:
L^\uparrow_+(\R^d)\to G(L)$, locality, spectral property and
cluster property.  Here $\tilde \tau$ is  obtained by analytic
 continuation of $\tau$
to a representation of the proper complex Lorentz group over $\C^d$
 (which contains $SO(d)$
as a real submanifold) and restriction of this representation to the
real orthochronous proper Lorentz group.
\end{theorem}
\noindent {\bf Proof.} The spectral property of $\hat
W_{n,(m_{l_1},j_1),\ldots , (m_{l_n},j_n)}^T$ is
a direct consequence of the
definition Eq. (\ref{7eqa}), cf. the proof of Proposition 7.8 of
\cite{AGW1}.

In order to prove the Laplace representation formula for
$S_{n,(m_{l_1},j_1),\ldots , (m_{l_n},j_n)}^T$
we proceed by induction over $\rho=\max\{j_l:l=1,\ldots,n\}$. The
statement for $\rho=1$ is just the statement of Proposition 7.8 in
\cite{AGW1}. Note that $(-\Delta+m^2)^{-j}(x)={(-1)\over (j-
1)}{d\over d m^2}(-\Delta+m^2)^{-(j-1)}(x)$ holds in the sense of
 tempered distributions. Let
$j_{a_1}=\ldots=j_{a_u}=\rho$ and all other $j_l<\rho$. For $j_l<\rho$
let $j_l'=j_l$ and $j_l'=j_l-1$ otherwhise. We then get by using Eq.
(\ref{6eqa}), the
induction hypothesis as well as continuity properties of the Laplace
transform and approximation of derivations by differential quotients:
\begin{eqnarray*}
S_{n,(m_{l_1},j_1),\ldots , (m_{l_n},j_n)}^T&=&\prod_{w=1}^u{(-1)
\over (j_{a_w}-
1)}{d\over d m_{j_{a_w}}^2}S_{n,(m_{l_1},j'_1),\ldots ,
(m_{l_n},j'_n)}^T\nonumber \\
&=& \prod_{w=1}^u{(-1)\over (j_{a_w}-
1)}{d\over d m_{j_{a_w}}^2}{\cal L}(\hat W_{n,(m_{l_1},j'_1),\ldots ,
(m_{l_n},j'_n)}^T)\nonumber \\
&=& {\cal L}(\prod_{w=1}^u{(-1)\over (j_{a_w}-
1)}{d\over d m_{j_{a_w}}^2}\hat W_{n,(m_{l_1},j'_1),\ldots ,
(m_{l_n},j'_n)}^T)\nonumber \\
&=& {\cal L}(\hat W_{n,(m_{l_1},j_1),\ldots ,
(m_{l_n},j_n)}^T),
\end{eqnarray*}
where we have made use of the explicit formulae (\ref{7eqa}),
 (\ref{7aeqa})  in the
last step. The above way of deriving the
$W_{n,(m_{l_1},j'_1),\ldots ,(m_{l_n},j'_n)}^T$ as
distributions w.r.t. some mass parameter can also be
 used to prove the
temperedness inductively, since we can differentiate terms
 like $\delta^{
\pm,(j)}_m$ and $(|k|^2+m^2)$  w.r.t. ${k^0}^2$ instead of
$m^2$ and use the fact that the change of variables
${k^0}^2\leftrightarrow k^0$ is smooth and polynomially bounded
for $k^2>\min\{ m_r:r=1\ldots,N\}-\epsilon>0$.

(ii) The Laplace representation of $S_{n,\alpha_1\cdots\alpha_n}^T$
immediately follows from (\ref{5eqa}), (\ref{6eqa}) and
the fact that ${\bf Q}_{E,n}(-i\underline{\nabla}){\cal L}(\hat W)
={\cal L}({\bf Q}_{M,n}(\underline{k})\hat W)$ for any
tempered distribution $\hat W$ on $\R^{dn}$ with the spectral property.
 The rest of the theorem
follows from \cite{OS} (for the cluster property, see the proof of
Theorem 7.10 of
\cite{AGW1})\kasten

By considerations similar to those proving the temperedness of the
$\hat W_n^T$ one can also show that the sequence of tempered
distributions fulfills the sufficient Hilbert space structure condition
on truncated Wightman functions introduced in \cite{AGW2}. Therefore,
the sequence of Wightman functions are the vacuum expectation values of
some quantum field theory in indefinite metric \cite{AGW2,MS}.

\section{Criteria for the existence of a scattering limit and
calculation of the scattering amplitudes}

In this section we replace the two point function $\hat W_2^T$ given
in Theorem 1 by a two point function $\hat {W'}_2^T(k_1,k_2)=
{\bf Q}_{M,2}(k_1,k_2)
\sum_{s=1}^N\lambda_s\delta^{-}_{m_s} (k_1)\delta (k_1+k_2)$,
$0\not = \lambda_s\in \R$ which clearly has the same covariance
 properties as
$\hat W_2^T$. The reason for this is that in any $\hat W_2^T$ there
occur terms $\hat W_{2,(m_{l_1},j_1)(m_{l_2},j_2)}^T$ with
$m_{l_1}=m_{l_2}$ and these terms lead to "exploding" scattering
behaviour, as we shall explain below. Replacing $\hat W_2^T$ by
$\hat{W'}_2^T$ simply means cancelling these terms in order to obtain
stable one-particle states. In some special cases this can be motivated
as a "renormalization procedure", see \cite{AIK}. We will also drop the
' and write $\hat W_2^T$ for the "new" 2-point functions from now on.

We now briefly recall some basic notions of axiomatic scattering theory
following \cite{He}.

For $l=1,\ldots,n$ let $\varphi_{l,s_l}^t(x)=(2\pi)^{-
dn/2}\int_{\R^d}e^{ik_l\cdot x}\hat \varphi_{l,s_l}(k) e^{i(k_l^0-
\omega_{l,s_l})t}~dk_l$, where $\hat \varphi_{l,s_l}$ is a
Schwartz function
with values in $\C^L$ and support in $\{ k_l^0>0,|k_l^2-
m_{s_l}^2|<\epsilon\}$ for $\epsilon$ small enough that non of these
neighbourhoods intersect for $m_{s_l}\not=m_{s'_l}$ and $\omega_{l,s_l}
=\sqrt{|k_l|^2+m_{s_l}^2}$.

We say that a sequence of truncated Wightman functions $W_n^T$ has
non trivial scattering behaviour, if the limits
\begin{eqnarray}
\label{9eqa}
&\displaystyle \langle \varphi_{1,s_1}\cdots\varphi_{r,s_r}^{\mbox{in}}|\varphi_{r+1,s_{r+1}}\cdots
\varphi_{n,s_n}^{\mbox{out}}\rangle^T=&\nonumber \\
& \displaystyle \lim_{t\to +\infty}
\int_{\R^{dn}} W^T_{n,\alpha_1\cdots\alpha_n}(x_1,\ldots,x_n)
\varphi^{\alpha_1, -t, *}_{r,s_r}(x_1)\cdots \varphi^{\alpha_r,-
t,*}_{1,s_1}(x_r)&\nonumber \\
&\displaystyle\times\varphi_{r+1,s_{r+1}}^{\alpha_{r+1},t}(x_{r+1})\cdots
\varphi_{n,s_n}^{\alpha_n,t}(x_n)dx_1\cdots dx_n; &\nonumber \\
&\displaystyle\langle \varphi_1\cdots\varphi_r^{\mbox{in/out}}|
\varphi_{r+1}\cdots \varphi_n^{\mbox{in/out}}\rangle^T
=&\nonumber\\
&\displaystyle \lim_{t\to \pm\infty}
\int_{\R^{dn}} W^T_{n,\alpha_1\cdots\alpha_n}(x_1,\ldots,x_n)
\varphi^{\alpha_1, -t ,*}_{r,s_r}(x_1)\cdots \varphi^{\alpha_r,-
t,*}_{1,s_1}(x_r)&\nonumber \\
&\displaystyle\times
\varphi_{r+1,s_{r+1}}^{\alpha_{r+1},-t}(x_{r+1})\cdots
\varphi_{n,s_n}^{\alpha_n, -t}(x_n)dx_1\cdots dx_n&
\end{eqnarray}
exist and are not identically equal to zero for $n\geq 3$. Here $*$ stands for
complex conjugation. By the following theorem we
give necessary and sufficient conditions, in terms of the mass spectrum
of $D$, for nontrivial scattering behaviour.

\begin{theorem}{Theorem}
(i) The sequence $W_n^T$ has nontrivial scattering behaviour, if and
only if $\nu_l=1$ for $l=1,\ldots,N$. If at least one of the $\nu_l$
 fulfills
$\nu_l>1$, then the first limit in Eq. (\ref{9eqa}) diverges
polynomially in $t$ for $n\geq 3$.

\noindent (ii) If $\nu_1=\cdots,\nu_N=1$, then
\begin{eqnarray}
\label{10eqa}
\langle \varphi_{1,s_1}^{\mbox{\rm ex}}|\varphi_{2,s_2}^{\mbox{\rm ex}}\rangle
^T &=& (2\pi)^{-d}\lambda_{s_1} \int_{\R^d\times\R^d}
\delta_{m_{s_1}}^+(k_1)\delta(k_1-k_2)
Q_{M,2}(k_1,k_2)_{\alpha,\beta}\nonumber \\
&&\times \hat \varphi_{1,s_1}^{\alpha, *}(k_1)\hat \varphi_{2,s_2}^{\beta
}(k_2) ~ dk_1dk_2 ~,
\end{eqnarray}
if $s_1=s_2$ and $\langle\varphi_{1,s_1}^{\mbox{\rm ex}}|
\varphi_{2,s_2}^{\mbox{\rm ex}}\rangle^T =0$
 otherwhise. Here, the two ${\rm ex}$ stand for all combinations of
$\rm in$ and
 $\rm out$. For $n\geq 3,$
\begin{eqnarray}
\label{11eqa}
&&\langle \varphi_{1,s_1}\cdots
\varphi_{r,s_r}^{\mbox{\rm in}}|\varphi_{r+1,s_{r+1}}\cdots
\varphi_{n,s_n}^{\mbox{\rm out}}\rangle^T  =\nonumber \\
&& (2\pi)^{(d(n-2)+4)/2} i \prod_{l=1}^Nb_{s_l,1} \int_{\R^{4n}}
 {\bf Q}_{M,n} (-k_1,\ldots,-
k_r,k_{r+1},\ldots,k_n)_{\alpha_1\cdots\alpha_n} \nonumber \\
&& \times \prod_{l=1}^r\delta_{m_{s_{r-l+1}}}^+(k_l) \prod_{l=r+1}^n
\delta_{m_{s_l}}^+(k_l)
\delta (\sum_{l=1}^r k_l-\sum_{l=r+1}^n k_l)\\
&& \times \hat\varphi_{r,s_r}^{\alpha_{1},*}(k_1)\cdots\hat\varphi_{1,s_1}^{
\alpha_{r}, *}(k_r)
 \hat \varphi_{r+1,s_{r+1}}^{\alpha_{r+1}}(k_{r+1})
\cdots\hat\varphi_{n,s_n}^{\alpha_n}(k_n) ~
dk_1\cdots dk_n \nonumber
\end{eqnarray}
and
\begin{equation}
\label{11aeqa}
\langle \varphi_{1,s_1}\cdots\varphi_{r,s_r}^{\mbox{\rm in}}|
\varphi_{r+1,s_{r+1}}\cdots \varphi_{n,s_n}^{\mbox{\rm in}}\rangle^T =
\langle \varphi_{1,s_1}\cdots\varphi_{r,s_r}^{\mbox{\rm out}}|
\varphi_{r+1,s_{r+1}}\cdots \varphi_{n,s_n}^{\mbox{\rm
out}}\rangle^T=0.
\end{equation}
\end{theorem}
\noindent {\bf Proof.} We first consider point (ii). This
is a rather straight forward generalization of the Theorems 1 and
2 of \cite{AGW3} and it can be obtained using the techniques developed
there, see \cite{Go} for the details of the proof and
further generalizations. Note that the condition that $Q_E(k)$ is prime
w.r.t the factors $(|k|^2+m_s^2)$ implies that ${\bf Q}_{M,n}$ does
not vanish identically on the mass-shells and therefore the scattering
amplitudes are not identically equal to zero.

It remains to prove point (i). Note that $({\partial\over \partial
{k^0_l}^2})=(\pm {1\over 2k^0_l}{\partial\over \partial k^0_l})$ on the support
of $\hat \varphi_{l,s_l}(\pm k)$. Therefore we get
$\delta^{\pm,(j_l-1)}_{m_{s_l}}(k_l)=(\pm {1\over 2k^0_l}{\partial\over
\partial k^0_l})^{j_l-1}\delta^\pm_{m_{s_l}}(k_l)$ and ${1\over (k_l^2-
m_{s_l}^2)^{j_l}}={(-1)^{j_l-1}\over (j_l-1)!}(\pm {1\over 2k^0_l}{\partial\over
\partial k^0_l})^{j_l-1}{1\over (k_l^2-m_{s_l}^2)}$ on these support sets. Using
Parseval's theorem and integration by parts we
get for the right hand side of Eq. (9)
$$
\int _{\R^{dn}}\hat W_{n,(m_r,1)\cdots(m_1,1)(m_{r+1},1)\cdots
(m_n,1)}^T(k_1,\ldots,k_n)\psi(t,k_1,\ldots ,k_n) ~ dk_1\cdots dk_n,
$$
where $\psi(t,k_1,\ldots,k_n)$ is a sum over functions of the type
$$\prod_{l=1}^n (\pm {1\over 2k^0_l}{\partial\over
\partial k^0_l})^{j_l-1}e^{\pm
i(k_l^0\pm\omega_{l,s_l})t}\hat h^{(*)}_{l}(\pm k_l),
$$
where $j_l$ takes values from $1$ to $\nu_l$ and the $\hat h_{l}$ are scalar functions
obtained by contraction of the $\hat \varphi^{\alpha_l}_{l,s_l}$ with
the tensor-valued polynomial ${\bf Q}_{M,n,\alpha_1,\ldots,\alpha_n}(\underline{k})$  and
multiplication with factors $1/k_l^0$ and
thus having the same support properties as the $\hat
\varphi^{\alpha_l}_{l,s_l}$. If $\nu_l>1$ for some $l$, such functions
can be written as a sum of functions of the type
$ t^r \times \prod_{l=1}^n e^{\pm i(k_l^0\pm\omega_{l,s_l})t}\hat g^{(*)}_{l}(\pm
k_l)$ with $0\leq r \leq \sum_{l=1}^n(\nu_l-1)$. If we consider the
unique term with maximal $r=\sum_{l=1}^n(\nu_l-1)\geq 1$, then, by (ii) of this theorem,
the corresponding term in the
scattering amplitude consists of an expression which converges to a
constant $\not = 0$ (provided the $\hat \varphi _{l,s_l}$
are chosen adequately) multiplied
by a factor $t^r$. Thus, the scattering amplitude in this case diverges polynomially
as $t\to \infty$.
\kasten

Theorem 2 shows that the scattering amplitudes
of some of the quantum field models diverge polynomially. This is a
remarkable fact, since it demonstrates that a generalization the
standard axiomatic scattering theory (Haag-Ruelle theory) for quantum fields  in positive
metric to the case of quantum fields in indefinite metric is
possible only under additional conditions (e.g. conditions on the infrared singularities
of the theory).

\

\noindent {\bf Aknowledgements.} We would like to thank C. Becker and R.
Gielerak for interesting discussions. The financial support of D.F.G.
via SFB 237 is gratefully acknowledged.

\end{document}